The experimental feature on the data of the primary proton identification in stratospheric X-ray emulsion chambers at energies >10 TeV (RUNJOB experiment).

Abstract.

The RUNJOB balloon-born emulsion chamber experiments have been carried out for investigating the composition and energy spectra of primary cosmic rays at energies 10-1000 TeV/nucleon. On the data of the treatment of RUNJOB` X-ray emulsion chambers exposed since 1995 to 1999 year about 50 % proton tracks were identified. In remained half of the events from proton group the single charged primary tracks were not found in the search area determined with high accuracy by the triangulation method using the several background heavy tracks. Considered methodical reasons in this paper could not explain this experimental result. The one from the probable physical reasons that is the neutrons in cosmic ray flux does not explain it too.

Introduction

The RUNJOB long duration balloon experiments were conducted since 1995 to 1999 year. The RUNJOB has 10 success long duration (6-7 days) flights from Kamchatka peninsula to Volga region. The total emulsion chamber exposure factor is 575[$m^2 \times$hour]. X-ray emulsion chambers (XEC) have been exposed at the average altitude about 32 km. The total statistics included to the resulting spectra is 521 events among them 360 protons [1]. The triangulation method using background heavy tracks closely located to the point of interaction alouds to search the nucleons and the nuclei of primary cosmic rays with high accuracy. For some events the accuracy of primaries location achieved a few ($\leq 10$) μm. According to result of the RUNJOB`95-99 XEC treatment for about 50% of proton events the single charged primary track was not found in the search circle with the radius ~ 3 $\sigma$, where $\sigma$ is location accuracy of primary particle.

With purpose of studying the methodical reasons to explain this experimental result the observed proton tracks of events from RUNJOB-6A, 5A chambers were traced on some emulsion layers above the interaction vertex. The result of this methodical work and the discussion of methodical questions concerning the proton identification in nuclear emulsions of XEC are presented below.

The experimental procedure of searching for the primary particles.

The nucleons and the nuclei of primary cosmic rays passing throw XEC (Fig.1) interacted with the substance of chamber and produced the secondary particles consisting from $\pi^{\pm}$, $\pi^0$ mesons mainly. Neutral pions initiated the electromagnetic cascades, which reveal themselves in the x-ray films as dark spots visible by the naked eye. During an exposure (6-7 days) there are more than 500 spots on lower x-ray film in the calorimeter.

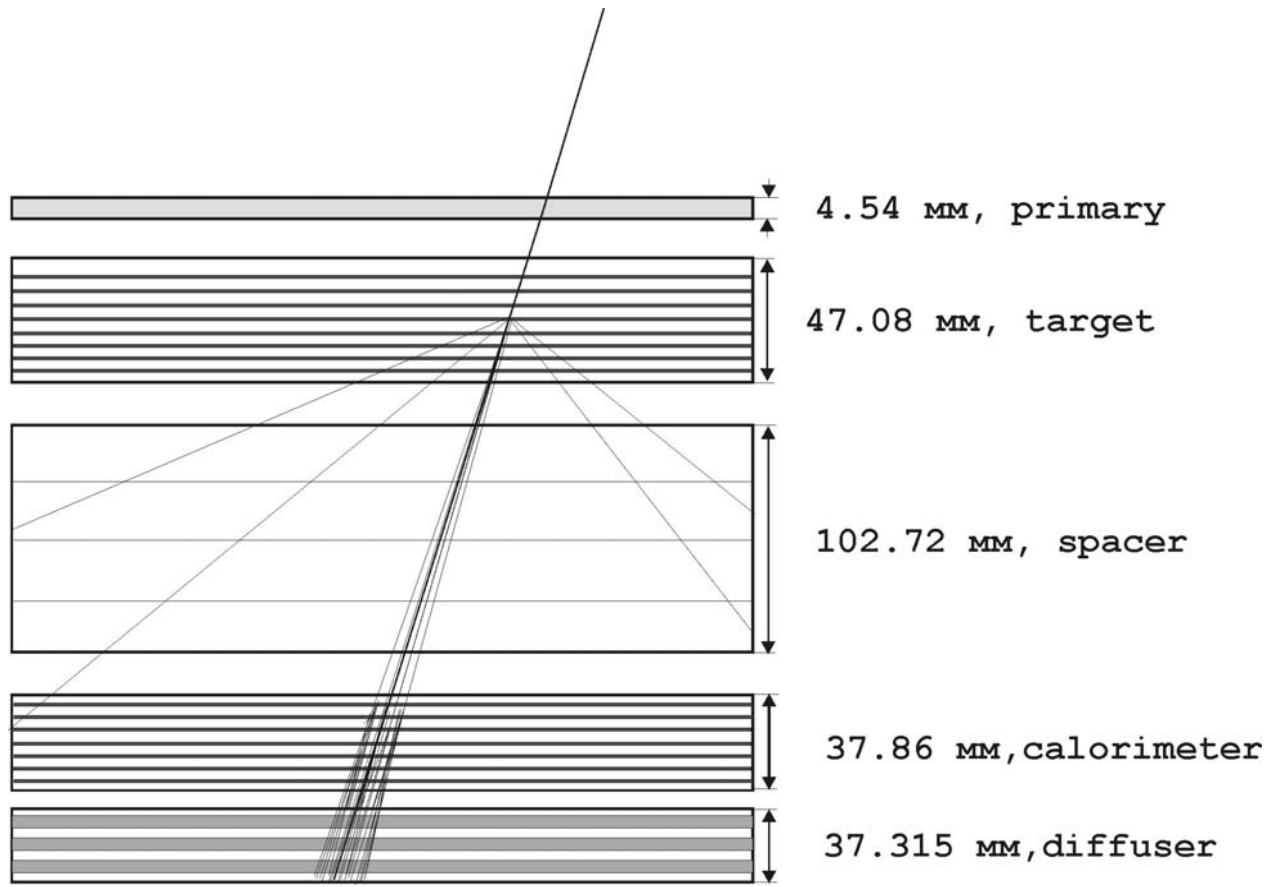

Fig.1 RUNJOB`97 Chamber

Events with x-ray film spots exceeding the selected threshold are located in an adjacent emulsion plates and then are traced in the upstream direction to the interaction vertex. In the emulsion plate above this vertex it was carried out the search of primary particle with the zenith and azimuth angles like ones of the secondary particle jet. During the long duration flights the emulsion plates have very high particle background. It was measured that there are 50÷70 tracks with different length and direction at exposure time $T_{exp} \sim 150$ h and on area $S=10^4$ $\mu m^2$. For definite identification of primaries the search area would be such that the condition was satisfied [2]: $N_{bg} \Delta S \Delta \Omega \ll 1$, where $N_{bg}$ is the number of the background tracks during the exposure time $T_{exp}$ of detector per unit of a solid angle on unit of area of detector; $\Delta S$ is accuracy of primary location; $\Delta \Omega = \sin\theta \Delta\varphi \Delta\theta$, where $\Delta\varphi$, $\Delta\theta$ are azimuth and zenith angle accuracies.

The next criteria must be satisfied for the primary particle candidate:
1. Measured zenith and azimuth angles of the particle candidate must be differed less $3^0$.
2. Ionization of particle must be about the same in all using layers.
3. The systematic trajectory deviation of the particle candidate must be not beyond the value of accuracy calculated shower trajectory.
4. The particle candidate ($Z \geq 2$) must be absent on layers in which the secondary particles are observed.

To decrease the primary particle location error in RUNJOB experiment it was used the triangulation method including the coordinate measurements of several background heavy tracks passing closely to each other (inside the circle with radius about 1÷2 cm). The primary particle location accuracy ($\sigma$) for individual events is from about 10 μm to about 150 μm.

The search area of primary particle is a circle with center in the location prediction point and with radius about $3\sigma$. In long duration experiments with XEC the accuracy $\sigma \sim 90 \div 200$ μm is enough for identification of primary nuclei with charge $\geq 6$ [3,4,5], but with this accuracy it is difficult to find the true small charged particle track among background. Decreasing of the location prediction error until $20 \div 50$ μm it is possible to identify helium tracks definitely. The further decrease of the search area (R < 60 mkm, $\sigma$ < 20 mkm) has to lead to the definite identification of proton. First of all the primary particle location accuracy depends on the event angle and the error of determination of the secondary particle emission center. According to data of the RUNJOB`97,99 chamber treatment for event groups with zenith angles tg $\theta \leq 1$ and tg $\theta > 1.7$ the average accuracies are $\sigma \cong 30$ mkm and $\sigma \cong 95$ mkm respectively. Fig2. shows distribution of found primary particle track coordinate deviations from predicted ones for events from RUNJOB-6A,11-A,B chambers with the energy threshold $\Sigma E_\gamma \geq 3$ TeV, $\Sigma E_\gamma \geq 5$ TeV, respectively, and with tg $\theta \leq 5$. The average value of deviations of measured coordinate from calculated ones equals: <$\Delta R$> = 57 mkm. For the comparison the location accuracy of other balloon-born emulsion chamber experiments (MUBEE, JACEE) is ~ 200 mkm and ~ 10 mkm, respectively. In RUNJOB experiment most of all primaries have been found in limits of $1\sigma$.

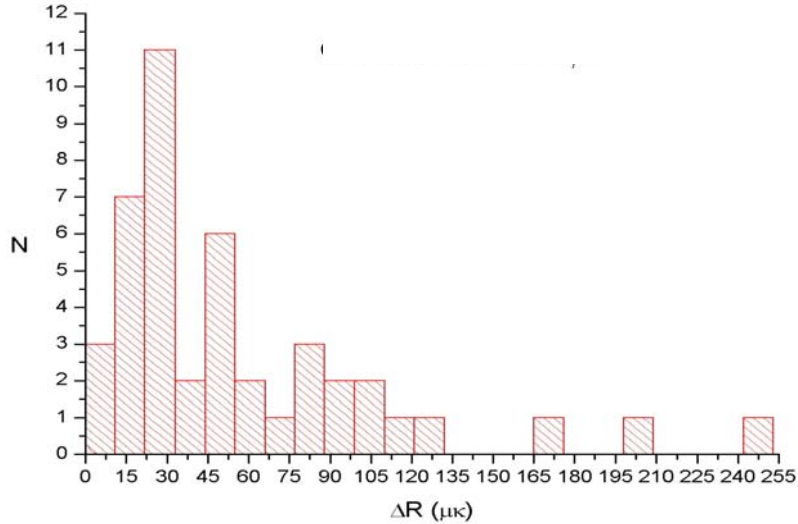

Fig.2 Deviations of measured coordinate of primary particles from calculated ones.

Registration of the protons in the RUNJOB` XEC

In process of the primary particle search as in RUNJOB experiment as in other balloon-born emulsion chamber experiments [3,4] it was considered that if the primary particle with charge $Z \geq 2$ is absent in the prediction region there should be proton (or may be neutron produced in interaction of primary particle in residual atmosphere).

In long duration balloon stratospheric flights with XEC on board the primary particle location accuracy ~10 mkm provides the definite identification of proton. In the experiment according to data of chamber treatments of RUNJOB`XEC the single charged track was found for $\approx 50\%$ proton events [3]. For example, after the primary particle search for the interactions with the energy threshold $\Sigma E_\gamma \geq 3$ TeV and $\Sigma E_\gamma \geq 5$ TeV and $\mathrm{tg}\theta \leq 5$ from the RUNJOB`97-6A and RUNJOB`99-11A,B chambers respectively twenty one events were identified as proton ones. Among them only for ten events the single charged track was found. In other cases (11 events) there is no primary candidate in the search area. Thus, it is needed to explain the reason for the absence of primary particle track for about half of events from proton group.

Let us look at the methodical reasons. Firstly, this is a bulky background particle complicating the identification of single charged tracks. However, secondary particles mainly pions are followed down in several layers, that is single charged particles are detected with about 100 % efficiency excepting cases when there is damages of emulsion. For example, characteristics of the events with measured secondary particle multiplicities are listed in Table 1.

| № event | A+B | $\theta^0$ | $N_{ch}$ | $\Sigma E_{ch}$ (TeV) | $\Sigma E_\gamma$ (TeV) | $E_0$ (TeV) |
|---|---|---|---|---|---|---|
| 9131140 (FIAN) [7] | ≤C + Pb | $13^0$ | 674 | 50 | 20 | 78 |
| 913281 (FIAN) [7] | O + Fe | $29^0$ | 239 | 12 | 15 | 30 |
| 913264 (FIAN) [7] | Li + Pb | $26^0$ | 290 | 40 | 20 | 67 |
| 40 (RUNJOB-IA) [8] | α + C | $11^0$ | 79 | | 3.4 | 9.5 |
| 45 (RUNJOB-IA) [8] | Mg + C | $27^0$ | 55 | 5.3 | | 8.2 |

Secondly, there is the accidental nature of error, for example the loss of particle at the search. To check this enough to search the primary in the emulsion some times or to trace the proton tracks on the second emulsion layer above the interaction vertex. If on the second layer the number of events with unfound proton tracks will be the same as on the first one we can conclude that the absence of the candidate one-charged tracks in layer above of the interaction vertex is explained by the accidental nature of error.

To analyze this error found protons from RUNJOB-6A, 5A chambers with energy threshold $\Sigma E_\gamma \geq 3$ TeV and $\Sigma E_\gamma \geq 5$ TeV, respectively, were followed further on some layers above interaction vertex. For the events with unfound primary tracks the location prediction was done ones more too. As a result we got:

1. Earlier found proton tracks were observed in two and more layers above the interaction vertex.
2. The new location prediction of primary track did not change the initial data of the existence or the absence of one-charged tracks in the search area.
3. For the one from eight events with unfound tracks the single charged track was found on the second layer above the interaction vertex.

It is desirable to mark the proton event with interaction vertex in base of the emulsion plate. The proton track was found in three layers above vertex in the limits of one standard deviation. It is necessary to note that the statistics of remeasured events is not sufficient for drawing conclusions about value of the random error. But this methodical work leads to the assumption that it is impossible to explain by methodical reasons only why the primary single charged tracks were not found for ~ 50% proton events from chambers exposed in 1995-1999 year.

A neutral component in flux of primary cosmic rays.

One could consider that the absence of single charged primary particle in the search area is explained by the neutral component of cosmic rays at the XEC exposure level.

If one suppose that these primaries are neutrons then it is difficult to explain such their amount since according to the calculations the relative amount of the high energy neutrons is small at observation level (~ 30 km). In paper [9] the relative contribution of the neutral secondaries with energies E > 20 TeV into total secondaries flux at the altitude about 10 g/cm$^2$ was presented. This contribution is 4.2 % in the angular interval $0^0$-$70^0$.

For the interactions from the proton event group detected in RUNJOB-6A и RUNJOB- 11A,B chambers with the energy threshold $\Sigma E_\gamma \geq$ 3 TeV и $\Sigma E_\gamma \geq$ 5 TeV respectively it was estimated the minimum multiplicity of secondary charged particle, that is the multiplicity without the count of particles in the central region under a assumption about the shape of forward region. In the proton events with unfound primary track the average minimum multiplicity $<N_{min}(-)>$ turned out about 1,5 times more then the middle minimal multiplicity in the proton events with observed tracks $< N_{min}(p)>$: $< N_{min}(-)> \approx$73, $< N_{min}(p)> \approx$ 40.

Possibly this difference suggests to the new kind of particles or the new type of interactions, although the statistics of the events is insufficient (21 events) for drawing conclusions.

Conclusions.

At the search of the primary particles with energies > 10 TeV/nucleon in RUNJOB XEC exposed in1995-1999 years for about 50% proton events the primary track was not found. This result it is difficult to explain by considerable particle background which nuclear emulsion has after long duration exposure since at first secondary single charged particles from interactions are detected with close to 100% efficiency and at second the error of the primary track location is sufficient for identification of proton and helium. On the average it is about 50 mkm for primaries with all angles.

On base of the treatment of events from RUNJOB-5A,6A chambers it would be possible to assume there is the random error of measurement but it is not enough to explain all amount of "proton" events with unfound the primary track.

On base of the calculations of cosmic ray passed throw residual atmosphere about 10 g/cm$^2$ it is impossible to explain that about 50% events with unfound primary tracks are the neutrons. If one suppose the observed interactions in XEC are induced by neutral particles then observed proton flux at energies > 10 TeV include considerable part of a neutral component. For drawing conclusion about the composition of the primary proton flux the future analysis of experimental material and may be future experiments are demanded.

References.